\documentclass[11pt,twoside]{article}

\usepackage{asp2006}
\usepackage{epsf}
\usepackage{psfig}
\usepackage{lscape}

\begin{document}
\title{Sustainability: A Tedious Path to Galactic Colonization}    %%% Fill in title
\author{Y. Dutil and S. Dumas}   %%% Fill in author names
\affil{Dept. de physique, de g\'enie physique et d'optique et Observatoire du mont M\'egantic, Universit\'e Laval, Qu\'ebec, Canada, G1K 7P4 }  %%% Fill in author affiliations

\begin{abstract} %%% Abstract to run on from here.

Civilization cannot sustain an exponential growth for long time even when neglecting numerous laws of physics! In this paper, we examine what are fundamental obstacles to long term survival of a civilization and its possibility to colonize the Galaxy. Using the solar system as a reference, resources available for sustained growth are analyzed. Using this information, we will explore the probability of discovering a civilization at its different stage of energy evolution as estimating some possible value of L, the typical life time of an extra-terrestrial civilization.

\end{abstract}

\section{Introduction}

Under the general assumption in SETI, extra-terrestrial civilization must be older than ours. However, little discussion has been made on how they should reach such a long lifetime.

\section{What physicists have to say?}

In 2010, our civilization will consume 17 TW of power (EIA 2006).  We might wonder what the physical limits to our power consumption are and what rate of growth is sustainable over time. We argue than the first growth crisis will be provoqued by climate change caused directly by energy production. We arbitrary fix this limit at 1 $W/m^2$ or 127 TW. This is slightly lower than the climate change produced by the increase of GHG on Earth (1.6 $W/m^2$, IPCC2007). With a 2\% growth rate, we will reach this limit within a century. This first step is highly critical since to grow further a civilization must have a complete control over all biophysical parameters of its ecosystem. 

The following steps can be described by the classical Kardashev civilization types. Keeping the growth rate at 2\%/yr, we would control all the light falling on Earth in 466 years (type I); in 1,533 years, we would control all the power output of our Sun (type II) and in 2,729 years, we would be able to control all the power of the Milky Way (level III)! Nevertheless, energetic considerations restrict the typical growth rate over a million years at a few ppm per year, which goes down to a few ppm over a billion years. In consequence, a civilization must stops growing very soon in its history (after a few hundred years at most). 

These physical limitations leave few possibilities for sustainable civilization. They could be photosynthesis limited ($\sim$ 10 TW), climatically limited (127 TW) or solar flux limited (174,000 TW).  It should be pointed out than the only fuel that can power a civilization over a billion years is deuterium. 

\section{What's economists have to say?}

The simplest economic model for growth has been proposed by Malthus in 1798. It simply states that population growth will be exponential while resources growth must be arithmetic which lead to a reduce wealth unless the population growth is stopped. Since then various economists have studied the population behavior and resources stocks for the case with renewable resources. These models are of the Ricardo-Malthus type (Ricardo 1817) and are a subtype of Lotka-Voltera predator-prey models \cite{lotka1925, Volterra1926}. These models produce three types of solutions: extinction, oscillation around a fixed point and stable steady states.

Such models have been successfully applied to population collapse of the Easter Island \cite{brander1998}. Generalizations of this model (Reuveny \& Decker 2000, Pezzey \& Anderies 2003) indicate than the {\em only escape of Malthusian trap is through institutions restricting utilization of resources}. This restriction itself is very difficult to implement effectively. 

Sustainability is even harder to achieve when non renewable resources are modeled. Some authors argue that technology will compensate for the natural capital loss \cite{Solow1997, Stiglitz1997} but others consider this as impossible \cite{GeorgescuRoegen1971, Daly1997}. Using a completely different approach (system dynamic analysis), Meadows et al (1972) came to the same conclusion.

Limitation of the action of the technology to insure sustainability has already being pointed out by the English economist Jeavons in 1865. Any technological amelioration leading to an improve efficiency will increase the affordability of this technology, which will then increase resources consumption. A modern formulation, known as the {\em Khazzoom-Brookes Postulate} \cite{brookes1990, khazzoom1980}, argues that energy saving innovations can end up causing even more energy to be used as the money saved is spent on other goods and services which themselves require energy in their production.

\section{What anthropologists have to say?}

Anthropologists have uncovered a rare example of strong sustainability for human civilization on a small pacific island: Tikopia. Against unfavorable odds, Tikopian have managed to survive on this isolated ecosystem for three millennia. Archaeological records show a first phase of sharp decline in forest areas, increased erosion, depletion of fish stocks and extinction of bird species, closely paralleling indigenous population growth. However, a striking finding by archaeologists is that the phase of environmental degradation was followed by a progressive historical change in society's resource-management practices.

Tikopian took effective measures between 1000 and 1800 AD to stabilize their population at approximately 1,281 to 1,323 people. They accomplished their goal by infanticide, abortion, and decreeing that only first-born sons could have children.  In addition, the inhabitants shifted from {\it slash and burn} practices to sustainable agriculture. Doing so, they have replaced the island natural ecosystem by an artificial one that mimics the structure and interrelationship found in natural ecologies. Finally, they eliminated pigs, despite the value Polynesians placed on them, because they damaged gardens and ate food than human could consume \cite{firth1983, kirsch2000}.

Amazingly, Tikopia sits on the cyclone belt, so every year its inhabitants deal with cyclones, every five or ten years, bringing heavy winds.  Not only Tikopia's society is sustainable but is also very resilient. 

\section{What ecologists have to say?}

Ecosystems are by definition sustainable. Selective pressures are hypothesized to drive evolution in one of two stereotyped directions: r- or K-selection \cite{macarthur1967, pianka1970}. These terms, r and K, are derived from standard ecological algebra, as illustrated in the simple Verhulst equation of population dynamics:
\begin{center}
\begin{displaymath} 
\frac{dN}{dt} = r N \left( 1 - \frac{N}{K} \right)
\end{displaymath}
\end{center}
where r is the growth rate of the population N, and K is the carrying capacity of its local environmental setting. Typically, r-selected species produce many offspring, which are, comparatively, less likely to survive to adulthood. Whereas K-selected species invest more heavily the nurture of fewer offspring, which has a better chance of surviving to adulthood.

In unstable or unpredictable environments r-selection predominates, where the ability to reproduce quickly is crucial, and there is little advantage in adaptations that permit successful competition with other organisms, because the environment is likely to change again. In stable or predictable environments K-selection predominates, as the ability to compete successfully for limited resources is crucial. Populations of K-selected organisms typically are very constant and close to the maximum that the environment can bear. It should be pointed out than in natural ecosystem biodiversity tend to increase both the stability and the productivity of ecosystem \cite{johnson1996, mcgrady1997}.

It is likely than any sustainable civilizations would follow a similar evolution trajectory. Therefore, alien civilizations are likely to be extremely complex, very efficient with a very low rate of growth.

\section{Conclusion}

A general conclusion is than strong sustainability can only be achieved by the implementation of strict control on the exploitation of resources. This control must be effectively achieved on a society which is likely to be extremely complex. The same society must also be very resilient to any large scale perturbation. These levels of social technologies are largely beyond our capabilities.

Since civilization collapse by resources exhaustion can happen on a very short time scale and social organization to avoid this is very difficult to put in place, it is likely that most civilization are short lived (few hundred years) and therefore unlikely to engage in galactic colonization. The lucky ones that are sustainable will have a zero growth rate, which would largely reduce the pressure for galactic colonization.

\end{document}